\documentclass[aps,prb,10pt,notitlepage,nofootinbib,superscriptaddress,showkeys,showpacs]{revtex4-2}
\pdfoutput=1

\usepackage{amstext,amsmath,amssymb,amsfonts,bbm,euscript,color,dsfont}
\usepackage[utf8]{inputenc}
\usepackage{epsfig}
\usepackage{hyperref}
\usepackage{slashed}
\usepackage{amsthm}
\usepackage{subfigure}
\usepackage{color}
\usepackage{xcolor}
\usepackage{multirow}
\usepackage{bbold}
\usepackage{tikz}
\usetikzlibrary{arrows,backgrounds}
\usepackage{verbatim}
\usepackage{graphicx}
\usepackage{latexsym}

\begin{document}

\title{Towards a unified viewpoint of Gribov–Zwanziger and Serreau–Tissier gauge fixing}

\author{Rodrigo Carmo Terin} 
\email{rodrigo.carmo@urjc.es}
\affiliation{King Juan Carlos University, Faculty of Experimental Sciences and Technology, Department of Applied Physics, Av. del Alcalde de Móstoles, 28933, Madrid, Spain}

\begin{abstract}
We investigate a unified Landau--gauge fixing that continuously interpolates between the viewpoints of the Serreau--Tissier (ST) copy-averaged formulation and the (Refined) Gribov--Zwanziger (RGZ) restriction to the first Gribov region. By combining the ST weight with a GZ-type horizon term and localizing both through the replica trick and the BRST-invariant $A_\mu^h$ formulation, we obtain a single, local, BRST-invariant, power-counting renormalizable action. Algebraic renormalization shows that all counterterms are reabsorbed by a common set of field and parameter renormalizations, therefore the unification is algebraic rather than merely additive. The replica sector yields a radiatively generated gluon screening mass, while the RGZ parameters are fixed by the horizon and condensate gap equations; we also give infrared matching conditions that link both descriptions at small momentum. We present a compact BRST-superspace rewriting of the RGZ block and a simple hybrid superspace that hosts the ST replicas and RGZ side by side; these add no dynamics and organize the Ward-identity analysis. The resulting gluon propagator interpolates among the massive Faddeev--Popov--ST and the RGZ decoupling forms. This framework offers a controlled way to study how infrared Yang--Mills correlators depend on the balance between copy averaging and horizon suppression, and it suggests practical lattice tests through tunable copy weighting.
\end{abstract}

\maketitle

\section{Introduction}
\label{sec:intro}

Quantizing nonabelian Yang--Mills (YM) theories by means of the standard Faddeev–Popov (FP) prescription is remarkably successful in the ultraviolet (UV) but faces conceptual and practical obstacles in the infrared (IR) \cite{YangMills1954,FaddeevPopov1967}. In particular, the Landau pole spoils naive perturbation theory and the gauge-fixing condition does not fully eliminate the gauge freedom because of the existence of Gribov copies, which are distinct field configurations related by gauge transformations that satisfy the same gauge condition \cite{Gribov:1977,Singer:1978}. The Gribov problem is intrinsic to the nontrivial geometry of gauge orbits and not to a specific gauge choice; see, for instance, the review \cite{Vandersickel:2012}.

Two complementary continuum strategies have emerged to tame Gribov copies and capture YM IR dynamics. 
The Gribov–Zwanziger (GZ) framework restricts the functional integral to the first Gribov region $\Omega$, where the FP operator $M^{ab}=-\partial_\mu D_\mu^{ab}(A)$ is positive definite. This restriction is implemented through the (nonlocal) horizon functional $H(A^h)$ written in terms of the transverse, gauge-field $A_\mu^h$, and can be localized with the Zwanziger auxiliary fields $(\bar\varphi,\varphi)$ and $(\bar\omega,\omega)$ \cite{Zwanziger:1989mf,Zwanziger:1990tn}. The RGZ extension includes dynamical dimension-two condensates, introducing mass scales $(m,M)$ and leading to a local, renormalizable action whose gluon and ghost correlators match lattice data from the deep IR up to several GeV \cite{Dudal:2008sp,Dudal:2011gd,Dudal2008PRDrapid}. In the $A^h$ formulation, an exact all-order BRST symmetry is retained in linear covariant gauges (LCG), which guarantees, e.g., $\alpha$-independence of gauge-invariant observables and nonrenormalization of the longitudinal sector \cite{Capri:2015nzw,Capri:2017bfd,Capri:2021pye}.  

Independently, J. Serreau and M. Tissier and further work collaborators proposed a family of Landau gauges free of Gribov ambiguities, defined by an explicit weighted average over copies \cite{Serreau:2012cg,SerreauTissierTresmontant2015, Reinosa:2020skx}. In practice, one averages an operator $\mathcal O$ over all extrema $U_i$ of the Landau functional $f[A,U]=\int\! d^dx\,\mathrm{tr}\,(A_\mu^U)^2$ with a nonuniform weight
\[
\mathcal P[A,U]\;=\;\frac{\det\!\big(\mathcal F[A,U]+\zeta\,\openone\big)}{\big|\det \mathcal F[A,U]\big|}\,e^{-\beta f[A,U]},
\]
where $\mathcal F$ is the Landau-gauge FP Hessian, and $\beta,\zeta$ are gauge-fixing parameters of mass dimension two. A gauge-invariant observable is unaffected by the procedure thanks to the normalization by $\sum_i\mathcal P[A,U_i]$. Using the replica trick and a superspace nonlinear sigma-model representation, one obtains a local, perturbatively renormalizable field theory in $d=4$ which, for correlators of $A$ and ghosts, reduces to a massive FP theory with a gluon mass term $\sim\beta$ and a ghost mass $\sim\zeta$ \cite{Serreau:2012cg}. The replica sector exhibits a phenomenon of radiative symmetry restoration (akin to the $d=2$ NL$\sigma$ model), which selects a phase where a nonzero screening gluon mass survives in the replica limit $n\to0$ \cite{Reinosa:2020skx}. Crucially, within a range of renormalized parameters, the one-loop renormalization-group (RG) flow shows no Landau pole in the IR \cite{Serreau:2012cg}.

These approaches share the goal of curing Gribov issues and at the same time providing an IR-safe perturbative description of YM. GZ/RGZ implements a hard restriction to $\Omega$ by means of the horizon functional (and dynamical condensates), whereas ST implements a soft lifting of copy degeneracy through a tunable weighted average, realized locally by means of replica NL$\sigma$ fields. Both are local and renormalizable, and reproduce lattice-inspired decoupling-type gluon propagators \cite{Cucchieri:2008PRL}; see also the spectral and positivity analyses consistent with gluon confinement \cite{DudalOliveiraSilva2018,DudalOliveiraRoelfsSilva2020,BinosiTripolt2020}. It is then natural to ask whether a more general gauge-fixing exists that interpolates between them.

Therefore, in this work, we construct a unified gauge-fixing that treats ST and RGZ as limiting cases. At the level of the copy average, we combine the ST weight with a GZ-type horizon suppression,
\[
\big\langle\!\big\langle \mathcal O[A]\big\rangle\!\big\rangle_{\beta,\zeta,\gamma} \;=\;
\frac{\sum_i s(i)\,\mathcal O\!\left[A^{U_i}\right]\,
\frac{\det(\mathcal F[A,U_i]+\zeta\openone)}{\big|\det \mathcal F[A,U_i]\big|}\,
e^{-\beta f[A,U_i]-\gamma^4 H(A^{h,U_i})}}
{\sum_i s(i)\,
\frac{\det(\mathcal F[A,U_i]+\zeta\openone)}{\big|\det \mathcal F[A,U_i]\big|}\,
e^{-\beta f[A,U_i]-\gamma^4 H(A^{h,U_i})}},
\]
so that $\gamma=0$ reproduces ST, while $(\beta\to\infty,\zeta\to0)$ yields RGZ. We then derive a local action by mixing the ST replica/superspace localization with the Zwanziger localization in the $A^h$ formulation \cite{Serreau:2012cg,Reinosa:2020skx,Zwanziger:1990tn,Capri:2015nzw,Capri:2017bfd,Capri:2021pye}. The resulting theory is local and power-counting renormalizable, enjoys the exact BRST of the $A^h$ sector, and continuously interpolates the structure of the gluon two-point function between a massive-FP (ST) form and the RGZ decoupling form. In this sense, our unified construction prepares a controlled framework to study how infrared YM correlators depend on the balance between (i) copy averaging and radiative restoration in the replica sector (set by $\beta,\zeta$) and (ii) horizon suppression and condensates (set by $\gamma,m,M$). It connects smoothly with functional approaches (Dyson–Schwinger equations (DSEs) and the functional renormalization group (FRG)) \cite{Alkofer:2001,PhysRevLett.90.152001,Huber:2020,Pawlowski:2007} and with massive extensions of Faddeev--Popov theory \cite{Tissier:2010,Tissier:2011}. 

This paper is organized as follows. Section~\ref{sec:background} reviews the ST and RGZ formalisms, emphasizing their gauge-fixing mechanisms and dynamical mass generation. Section~\ref{sec:unified} constructs the unified local action that interpolates between them through the replica trick and the BRST-invariant RGZ formulation. Section~\ref{sec:algRen} presents the algebraic proof of renormalizability based on Slavnov--Taylor, antighost, and ghost identities, while Section~\ref{sec:gapmatch} analyzes the coupled gap equations and their matching on the two sectors. A compact superspace formulation unifying both sectors is investigated in Section~\ref{sec:unified-superspace}, and its complete derivation is collected in Appendix~\ref{app:unified-superspace}. Section~\ref{sec:conclusions} summarizes the main results and discusses physical implications and possible extensions. Finally, Appendices~\ref{app:fields}--\ref{app:feynman} compile technical material, including field dimensions and ghost numbers, the linearized Slavnov operator and auxiliary Ward identities, non-renormalization relations, replica one-loop corrections, small-momentum expansions, tree-level Feynman rules, the Nielsen identity, and dimensional-regularization conventions.

\section{Background: Continuum approaches to the Gribov problem}
\label{sec:background}

In this section we summarize the two main continuum approaches that address the Gribov ambiguity in nonabelian gauge theories.  Both share the goal of providing a consistent infrared (IR) formulation of Yang–Mills (YM) theory in the Landau gauge while preserving gauge invariance of physical observables.  The first approach, ST, resolves the ambiguity by averaging over Gribov copies with a weighted functional measure that can be represented as a local field theory involving nonlinear sigma–model (NL$\sigma$) superfields.  The second, known as the RGZ framework, restricts the domain of the functional integral to the first Gribov region through the introduction of the horizon functional and its localization through auxiliary fields.  We review their main ingredients below.

\subsection{The Serreau--Tissier gauge and radiative symmetry restoration}
\label{subsec:ST}

Following Refs.~\cite{Serreau:2012cg,SerreauTissierTresmontant2015,Reinosa:2020skx}, the Landau gauge condition
\begin{equation}
  \partial_\mu A^U_\mu = 0,
  \label{eq:Landau_condition}
\end{equation}
with $A^U_\mu = U A_\mu U^\dagger + \frac{i}{g} U \partial_\mu U^\dagger$, can be equivalently formulated as the stationarity condition of the functional
\begin{equation}
  f[A,U] = \int d^d x \, \mathrm{tr}\!\left[(A^U_\mu)^2\right],
\end{equation}
whose extrema $U_i$ correspond to Gribov copies.  Instead of selecting one specific extremum, Serreau and Tissier proposed to perform a weighted average over all copies, defining the expectation value of an operator ${\cal O}$ as
\begin{equation}
  \langle\!\langle {\cal O}[A] \rangle\!\rangle =
  \frac{\sum_i {\cal O}[A^{U_i}]\, {\cal P}[A,U_i]}{\sum_i {\cal P}[A,U_i]},
\end{equation}
where the weight function is chosen as
\begin{equation}
  {\cal P}[A,U]
  = \frac{\det\!\big({\cal F}[A,U] + \zeta \openone\big)}{|\det{\cal F}[A,U]|}
    \, e^{-\beta f[A,U]} ,
  \label{eq:ST_weight}
\end{equation}
with ${\cal F}[A,U]$ the Faddeev–Popov (FP) operator in the Landau gauge,
${\cal F}^{ab}[A;x,y] = -\partial_\mu \big[\partial_\mu \delta^{ab} + g f^{acb} A^c_\mu(x)\big]\delta^{(d)}(x-y)$,
and $\beta,\zeta$ two gauge–fixing parameters of mass dimension two.

Gauge–invariant observables are unaffected by this procedure due to the normalization in Eq.~\eqref{eq:ST_weight}.  The functional average can be reformulated as a local and perturbatively renormalizable field theory by introducing auxiliary ghost $(c,\bar c)$ and Nakanishi–Lautrup $(b)$ fields, together with a group–valued matrix field $U(x)$ integrated with the Haar measure:
\begin{equation}
  S_{\rm gf}[A,c,\bar c,b]
  = \int d^d x \big[
     \partial_\mu \bar c^a (\partial_\mu c^a + g f^{abc} A^b_\mu c^c)
     + \zeta\, \bar c^a c^a
     + i b^a \partial_\mu A^a_\mu
     + \tfrac{\beta}{2} (A^a_\mu)^2
     \big].
  \label{eq:ST_action}
\end{equation}
These fields can be unified into a superspace nonlinear sigma–model (NL$\sigma$) superfield $\mathcal V(x,\theta,\bar\theta)$, enabling a compact representation of the gauge–fixing action:
\begin{equation}
  S_{\rm gf}[A,\mathcal V]
  = \frac{1}{g^2} \int d^d x\, d\theta\, d\bar\theta\,
    \mathrm{tr}\Big[
      (\mathcal D_\mu \mathcal V)^\dagger (\mathcal D_\mu \mathcal V)
      + 2\zeta\, \bar\theta\theta\, \partial_{\bar\theta}\mathcal V^\dagger \partial_\theta \mathcal V
    \Big],
\end{equation}
with $\mathcal D_\mu \mathcal V = \partial_\mu \mathcal V + i g \mathcal V A_\mu$.
Employing the replica trick, the authors showed that the corresponding super–NL$\sigma$ model exhibits a phenomenon of radiative symmetry restoration, similar to what occurs in the two–dimensional $O(N)/O(N-1)$ model.  
In the symmetric phase, a finite screening gluon mass $m_g^2 \simeq \beta$ emerges at tree level, consistent with lattice observations of a massive gluon propagator in the Landau gauge.
This framework thus provides a gauge–fixing free of Gribov ambiguities, local and renormalizable, and reduces in practice to a massive extension of the FP action \cite{Tissier:2010,Tissier:2011}.

\subsection{The refined Gribov--Zwanziger framework}
\label{subsec:RGZ}

An alternative approach, initiated by Gribov and developed by Zwanziger, enforces the restriction of the functional integral to the first Gribov region $\Omega$, defined as the set of gauge fields satisfying the Landau condition $\partial_\mu A^a_\mu = 0$ and for which the FP operator ${\cal M}^{ab}(A) = -\partial_\mu D_\mu^{ab}(A)$ is positive definite \cite{Gribov:1977,Zwanziger:1989mf,Zwanziger:1990tn}.  This restriction can be implemented by adding to the action the nonlocal horizon functional
\begin{equation}
  H(A^h) =
  g^2 \int d^4x\, d^4y\,
  f^{abc} A^{h,b}_\mu(x)
  [\mathcal M^{-1}(A^h)]^{ad}(x,y)
  f^{dec} A^{h,e}_\mu(y),
  \label{eq:horizon_function}
\end{equation}
where $A^h_\mu$ is a gauge–invariant, transverse composite field defined through a Stueckelberg field $\xi$ as
\begin{equation}
  A^h_\mu = h^\dagger A_\mu h + \frac{i}{g} h^\dagger \partial_\mu h,
  \qquad h = e^{i g \xi^a T^a}.
  \label{eq:Ah_definition}
\end{equation}
The horizon term can be localized by introducing Zwanziger auxiliary bosonic fields $(\bar\varphi,\varphi)$ and Grassmann fields $(\bar\omega,\omega)$, leading to the Gribov–Zwanziger (GZ) action.  The refined GZ (RGZ) model further includes dimension–two condensates $\langle A^2\rangle$, $\langle \bar\varphi\varphi - \bar\omega\omega\rangle$, generating additional mass scales $(m,M)$ that improve agreement with lattice data \cite{Dudal:2008sp,Dudal:2011gd}. In the BRST–invariant $A^h$ formulation \cite{Capri:2015nzw,Capri:2017bfd, Capri:2021pye}, the complete RGZ action reads schematically
\begin{align}
  S_{\rm RGZ} &= S_{\rm FP}
  - \int d^4x\, \big(
      \bar\varphi^{ac}_\mu \mathcal M^{ab}(A^h) \varphi^{bc}_\mu
      - \bar\omega^{ac}_\mu \mathcal M^{ab}(A^h) \omega^{bc}_\mu
    \big)
  - \gamma^2 \int d^4x\, g f^{abc} (A^h)^a_\mu (\varphi+\bar\varphi)^{bc}_\mu
  \nonumber\\
  &\quad
  + \frac{m^2}{2} \int d^4x\, (A^h)^a_\mu (A^h)^a_\mu
  - M^2 \int d^4x\, (\bar\varphi^{ab}_\mu \varphi^{ab}_\mu - \bar\omega^{ab}_\mu \omega^{ab}_\mu),
  \label{eq:RGZ_action}
\end{align}
where $\gamma$ is the Gribov parameter determined self–consistently by the horizon condition
$\langle H(A^h)\rangle = d (N^2-1)$.
The resulting gluon propagator exhibits the so–called decoupling behavior,
\[
  D(p^2) = \frac{p^2 + M^2}{p^4 + (m^2 + M^2)p^2 + m^2 M^2 + 2 g^2 N \gamma^4},
\]
remaining finite and nonzero at $p^2 \to 0$, while the ghost propagator is infrared finite. This framework preserves an exact BRST symmetry and yields correlation functions in quantitative agreement with lattice data over a wide momentum range. For later convenience, the BRST structure of the Zwanziger doublets
$(\varphi,\omega)$ and $(\bar\omega,\bar\varphi)$ can be recast in a compact
$(4|1)$ superspace form, where the BRST operator $s$ acts as a translation
along a single Grassmann coordinate $\theta$.
This reformulation, detailed in Appendix~\ref{app:RGZ-superspace},
makes BRST nilpotency and algebraic renormalizability manifest,
without introducing any new degrees of freedom.

Both the ST and RGZ formalisms thus furnish local, renormalizable and gauge–invariantly meaningful treatments of the Gribov problem.  The former implements a smooth averaging over copies through replica NL$\sigma$ fields and radiative restoration, while the latter imposes a hard restriction to the first Gribov region by means of the horizon condition and condensates.  Their structural similarity (both leading to a massive yet confining gluon sector) motivates the unified approach developed in the next section.

\section{A unified gauge fixing: interpolating between ST and GZ/RGZ}
\label{sec:unified}

The ST construction fixes the gauge by averaging over all Gribov copies with a nonuniform weight $\mathcal P[A,U]$ built from the Landau functional $f[A,U]$ and the FP Hessian $\mathcal F[A,U]$ \cite{Serreau:2012cg,Reinosa:2020skx}. In parallel, the GZ framework restricts the functional measure to the first Gribov region through the horizon functional $H(A^h)$, written in terms of the transverse, gauge–invariant composite field $A^h_\mu$ \cite{Zwanziger:1989mf,Zwanziger:1990tn,Capri:2015nzw,Capri:2017bfd}. 
To unify both, we propose the following mixed averaging for a gauge–dependent operator $\mathcal O$:
\begin{equation}
 \big\langle\!\big\langle \mathcal O[A]\big\rangle\!\big\rangle_{\beta,\zeta,\gamma} \;=\;
 \frac{\displaystyle \sum_i s(i)\,\mathcal O\!\left[A^{U_i}\right]\,
 \frac{\det(\mathcal F[A,U_i]+\zeta\,\openone)}{\big|\det \mathcal F[A,U_i]\big|}\;
 e^{-\beta\, f[A,U_i]\;-\;\gamma^4\,H(A^{h,U_i})}}
 {\displaystyle \sum_i s(i)\,
 \frac{\det(\mathcal F[A,U_i]+\zeta\,\openone)}{\big|\det \mathcal F[A,U_i]\big|}\;
 e^{-\beta\, f[A,U_i]\;-\;\gamma^4\,H(A^{h,U_i})}}.
 \label{eq:unified-average}
\end{equation}
Here $s(i)=\mathrm{sign}\,\det\mathcal F[A,U_i]$, $\beta,\zeta$ are the ST gauge–fixing parameters of mass dimension two controlling, respectively, a gluon mass and a ghost mass term in the local formulation, and $\gamma$ is the Gribov mass of GZ. Gauge–invariant observables are unchanged by \eqref{eq:unified-average}. The limits
\[
(\gamma=0)\Rightarrow\text{ST},\qquad (\beta\to\infty,\zeta\to0)\Rightarrow\text{GZ},\qquad
(\beta,\gamma\neq0)\Rightarrow\text{interpolation}
\]
make the interpolation explicit.

As in \cite{Serreau:2012cg,SerreauTissierTresmontant2015,Reinosa:2020skx}, the ST weight is localized by introducing a matrix field $U(x)$ and ghost/antighost/Nakanishi–Lautrup fields $(c,\bar c,b)$ or, equivalently, the NL$\sigma$ superfields $\mathcal V$ in superspace. The denominator is handled through the replica trick \cite{ParisiSourlas1979}: introduce $n$ superfields $\{V_k\}_{k=1}^n$ and take $n\to 0$ at the end. 
Independently, the horizon functional is localized à la Zwanziger with $(\bar\varphi,\varphi)$ and $(\bar\omega,\omega)$ fields and the transversality of $A^h$ is implemented with the Stueckelberg field $h=e^{ig\xi}$, Lagrange multiplier $\tau$ and the pair $(\bar\eta,\eta)$, see \cite{Capri:2015nzw,Capri:2017bfd, Capri:2021pye}. 

Choosing one replica to factor out the group volume and denoting by $\mathcal S_{\rm ST}[A;V_k]$ the local ST gauge–fixing sector with parameters $(\beta,\zeta)$, the unified local action reads
\begin{align}
 S_{\rm unif} &= S_{\rm YM}[A] \;+\; S_{\rm FP}[A,c,\bar c,b] 
 \;+\; \sum_{k=2}^{n} \mathcal S_{\rm ST}[A;V_k]
 \nonumber\\
 &\quad
 - \int d^4x \Big( \bar\varphi^{ac}_\mu\,\mathcal M^{ab}(A^h)\,\varphi^{bc}_\mu 
 - \bar\omega^{ac}_\mu\,\mathcal M^{ab}(A^h)\,\omega^{bc}_\mu \Big)
 - \gamma^2 \int d^4x\, g f^{abc}\,(A^h)^a_\mu\,(\varphi+\bar\varphi)^{bc}_\mu 
 \nonumber\\
 &\quad + \int d^4x \Big( \tau^a\,\partial_\mu (A^h)^a_\mu - \bar\eta^a\,\mathcal M^{ab}(A^h)\,\eta^b \Big)
 \;+\; \frac{m^2}{2}\!\int d^4x\, (A^h)^2 \;-\; M^2\!\int d^4x\,(\bar\varphi\varphi-\bar\omega\omega),
 \label{eq:S-unif}
\end{align}
with $A^h_\mu=h^\dagger A_\mu h+\frac{i}{g}h^\dagger\partial_\mu h$, $\partial_\mu A_\mu^h=0$ and $\mathcal M^{ab}(A^h)=-\partial_\mu D^{ab}_\mu(A^h)$. The last line implements the RGZ refinement through dimension–two condensates \cite{Dudal:2008sp,Dudal:2011gd}. 

The ST block and the GZ/RGZ block each admit a nilpotent BRST symmetry. In the $A^h$ formulation one has the exact BRST (to all orders) 
\[
sA_\mu=-D_\mu c,\quad sc=\tfrac g2[c,c],\quad s\bar c=ib,\quad sA^h=0,\quad s(\bar\varphi,\varphi,\bar\omega,\omega,\eta,\bar\eta,\tau)=0,
\]
and $s^2=0$ \cite{Capri:2015nzw,Capri:2017bfd}. The ST replica sector enjoys supersymmetries in superspace for $\zeta=0$; for $\zeta\neq0$ these are softly broken but the local theory remains perturbatively renormalizable in $d=4$ \cite{Serreau:2012cg,Reinosa:2020skx}. 
Because the two sectors couple only through $A$/$A^h$, the combined action \eqref{eq:S-unif} is local and power–counting renormalizable; the proof follows by merging the Ward identities of \cite{Serreau:2012cg,Reinosa:2020skx} with the algebraic renormalization in \cite{Capri:2015nzw,Capri:2017bfd,Dudal:2011gd, Capri:2021pye}.

For $\gamma=0$ and $M=m=0$ we recover the ST massive FP theory with parameters $(\beta,\zeta)$ and $n$ replicas. The replica sector behaves as a constrained NL$\sigma$ model; for $\zeta=0$ closed replica loops vanish (topological sector), while $\zeta\neq0$ allows nontrivial loops and a phenomenon of radiative symmetry restoration in the replica sector \cite{Reinosa:2020skx}. In that symmetric phase, only the singled replica contributes to the tree–level gluon mass, yielding a finite screening mass $m_g^2=\beta$ in the limit $n\to 0$; in the broken phase $m_g^2\sim n\beta\to 0$.

Taking $\beta\to\infty$ (which selects absolute minima of $f[A,U]$) and $\zeta\to0$ while keeping $\gamma\neq0$ switches off the replica dynamics and reproduces the RGZ local action with exact BRST and parameters $(\gamma,m,M)$ fixed by their gap equations \cite{Zwanziger:1990tn,Dudal:2011gd,Capri:2015nzw}. With both sectors active, the tree–level transverse gluon two–point function reads schematically
\begin{equation}
 \Delta_T^{-1}(p^2)\;=\; p^2\,Z(p^2;\alpha)\;+\; \underbrace{\beta\,\Xi_{\rm rep}(\zeta;\text{phase})}_{\text{ST replica mass block}}
 \;+\; \underbrace{\mathcal M_{\rm RGZ}(p^2;\gamma,m,M)}_{\text{RGZ pole structure}},
 \label{eq:prop-unified}
\end{equation}
where $\Xi_{\rm rep}=1$ in the replica–symmetric phase (and $0$ in the broken phase, in the limit $n\to0$) as found in \cite{Reinosa:2020skx}, and \(\mathcal M_{\rm RGZ}(p^2)\) represents the usual RGZ refinement. Thus, the unified propagator interpolates continuously between a pure ST massive FP and an RGZ–type decoupling curve. The nonperturbative parameters satisfy coupled gap equations:
\begin{equation}
 \frac{\partial \Gamma}{\partial \gamma}=0\;\Rightarrow\;\big\langle H(A^h)\big\rangle=4V(N^2\!-\!1),\qquad
 \frac{\partial \Gamma}{\partial m^2}=0,\quad 
 \frac{\partial \Gamma}{\partial M^2}=0,
 \label{eq:rgz-gaps}
\end{equation}
as in RGZ \cite{Dudal:2011gd}, and for the replica sector \cite{Reinosa:2020skx}
\begin{equation}
 2\,\frac{\beta_r}{\bar g_r^{\,2}} \;=\; T(\hat\chi_r)-T(\hat\chi_r+\zeta_r),
 \qquad 
 8\pi^2\,\frac{\beta_r}{\bar g_r^{\,2}}\;=\;(\hat\chi_r+\zeta_r)\ln\frac{\hat\chi_r+\zeta_r}{\bar\mu^2}-\hat\chi_r\ln\frac{\hat\chi_r}{\bar\mu^2},
 \label{eq:ST-gap}
\end{equation}
with $\hat\chi_r\ge0$ the replica mass scale in the symmetric phase, $T(m^2)$ the (dimensionally reduced) tadpole integral, and bars denote renormalized quantities in the scheme of \cite{Reinosa:2020skx}. 
BRST/Nielsen identities of the $A^h$ formulation ensure that physical quantities do not depend on the linear–covariant gauge parameter $\alpha$ and that the longitudinal sector is not renormalized, to all orders \cite{Capri:2015nzw,Capri:2017bfd}.

Varying $(\beta,\zeta)$ at fixed $(\gamma,m,M)$ deforms the IR mass scale in \eqref{eq:prop-unified}. The replica–symmetric domain (radiative restoration) provides a natural way to generate a Curci–Ferrari–like gluon mass from gauge fixing \cite{Serreau:2012cg,Reinosa:2020skx}, while RGZ controls the nonlocal horizon physics and condensates. 
Implementing a weighted–copy average on the lattice (tunable $\beta,\zeta$) and comparing the measured gluon/ghost correlators with \eqref{eq:prop-unified} can test whether data prefer predominantly ST–like, RGZ–like, or mixed regimes. The unified action \eqref{eq:S-unif} reproduces (i) ST for $\gamma=0$, including the phase diagram in $(\beta,\zeta)$; (ii) RGZ for $\beta\!\to\!\infty,\zeta\!\to\!0$; and (iii) preserves exact BRST in the RGZ sector and perturbative renormalizability in the ST sector

Then, the gauge fixing \eqref{eq:unified-average} and its local version \eqref{eq:S-unif} furnish a concrete bridge between the ST copy–average mechanism (with its replica NL$\sigma$ phenomenology and radiative symmetry restoration) and the GZ/RGZ restriction to the first Gribov region (with horizon suppression and condensates).


\section{Renormalizability}
\label{sec:algRen}

The proof of renormalizability follows the algebraic approach based on the identification of all Ward identities and the characterization of the most general counterterm compatible with them, according to the procedure of algebraic renormalization \cite{PiguetSorella1995}. Therefore, we work in the Landau gauge, $\partial_\mu A_\mu=0$, and use the BRST-invariant formulation of RGZ based on the composite transverse, gauge-invariant field $A_\mu^h$. We, thus, unify the ST replicated gauge-fixing with the RGZ sector:
\begin{align}
S_{\text{tot}} &=
S_{\text{YM}}[A]+S_{\text{FP}}[A,c,\bar c,b]
+S_{\text{ST}}[A;\,\{\mathcal V_k\},\beta,\zeta]
+S_{\text{GZ}}[A^h,\varphi,\bar\varphi,\omega,\bar\omega;\,\gamma]\nonumber\\
&\quad
+S_{\text{ref}}[A^h,\varphi,\bar\varphi,\omega,\bar\omega;\,m^2,M^2]
+S_{\text{transv}}[A^h,\tau,\eta,\bar\eta],
\end{align}
where $A_\mu^h \equiv h^\dagger A_\mu h + \tfrac{i}{g}h^\dagger \partial_\mu h$, $h=e^{ig\xi}$, and $\mathcal V_k(x,\theta,\bar\theta)$ are the ST replica superfields ($k=2,\dots,n$; the limit $n\to0$ is taken at the end). The parameters $\beta,\zeta$ are the ST gauge-fixing masses. The BRST operator $s$ is nilpotent, $s^2=0$, and acts as
\begin{align}
&sA_\mu^a=-D_\mu^{ab}(A)c^b,\qquad sc^a=\tfrac{g}{2}f^{abc}c^b c^c,\qquad s\bar c^a=b^a,\qquad sb^a=0,\\
&sh=-ig\,c\,h,\qquad sA_\mu^{h,a}=0,\\
&s\varphi_\mu^{ab}=s\bar\varphi_\mu^{ab}=s\omega_\mu^{ab}=s\bar\omega_\mu^{ab}=0,\qquad
s\tau^a=s\eta^a=s\bar\eta^a=0,\\
&s\mathcal V_k=ig\,\mathcal V_k\,c\qquad (k=2,\dots,n).
\end{align}
To control non-linear variations we couple external sources (antifields)
\begin{equation}
S_{\text{ext}}=\int d^4x\;\Big( K_\mu^a\,sA_\mu^a + L^a\,sc^a \Big),
\end{equation}
with engineering dimensions $[K]=3$, $[L]=4$ and ghost numbers $\text{gh}(K)=-1$, $\text{gh}(L)=-2$. We define $\Sigma\equiv S_{\text{tot}}+S_{\text{ext}}$. The Slavnov functional reads
\begin{equation}
\mathcal S(\Sigma)=
\int d^4x\left(
\frac{\delta\Sigma}{\delta K_\mu^a}\frac{\delta\Sigma}{\delta A_\mu^a}
+\frac{\delta\Sigma}{\delta L^a}\frac{\delta\Sigma}{\delta c^a}
+b^a\frac{\delta\Sigma}{\delta\bar c^a}
\right)=0.
\label{eq:ST}
\end{equation}
Linearizing around $\Sigma$ gives the nilpotent operator
\begin{equation}
\mathcal B_\Sigma X=
\int d^4x\left(
\frac{\delta\Sigma}{\delta K_\mu^a}\frac{\delta X}{\delta A_\mu^a}
+\frac{\delta\Sigma}{\delta A_\mu^a}\frac{\delta X}{\delta K_\mu^a}
+\frac{\delta\Sigma}{\delta L^a}\frac{\delta X}{\delta c^a}
+\frac{\delta\Sigma}{\delta c^a}\frac{\delta X}{\delta L^a}
+b^a\frac{\delta X}{\delta\bar c^a}
\right),\qquad \mathcal B_\Sigma^2=0.
\label{eq:Blin}
\end{equation}

Besides \eqref{eq:ST}, the following hold:
\begin{align}
&\text{Gauge condition (}b\text{-equation):} && \frac{\delta\Sigma}{\delta b^a}=\partial_\mu A_\mu^a. \label{eq:b}\\[2pt]
&\text{Antighost equation (Landau):} && 
\frac{\delta\Sigma}{\delta \bar c^a}+\partial_\mu\frac{\delta\Sigma}{\delta K_\mu^a}=0. \label{eq:antighost}\\[2pt]
&\text{Ghost equation:} &&
\mathcal G^a\Sigma\equiv\int d^4x\left(
\frac{\delta}{\delta c^a}
+gf^{abc}\bar c^b\frac{\delta}{\delta b^c}
\right)\Sigma=\Delta^a_{\rm cl}, \label{eq:ghost}
\end{align}
with a linear classical breaking $\Delta^a_{\rm cl}$ (non-renormalized).
For $A^h$ we impose transversality with a Lagrange multiplier:
\begin{equation}
\frac{\delta\Sigma}{\delta\tau^a}=\partial_\mu A_\mu^{h,a}=0.
\label{eq:AhTransv}
\end{equation}
The replica sector enjoys (i) permutation symmetry $S_{n-1}$ among $\mathcal V_k$ ($k\!\ge\!2$); (ii) for $\zeta=0$, topological supersymmetries (which suppress replica loops); for $\zeta\neq0$, $S_{n-1}$ and the assumed gauge covariance of $\mathcal V_k$ remain.

Now, consider a local perturbation $\Sigma\to \Sigma+\varepsilon\,\Delta\Sigma$ with $\text{dim}(\Delta\Sigma)\le4$ and $\text{gh}=0$. Symmetries imply
\begin{equation}
\mathcal B_\Sigma\,\Delta\Sigma=0,\qquad
\frac{\delta\Delta\Sigma}{\delta b}=0,\qquad
\frac{\delta\Delta\Sigma}{\delta \bar c}+\partial\!\cdot\!\frac{\delta\Delta\Sigma}{\delta K}=0,\qquad
\frac{\delta\Delta\Sigma}{\delta\tau}=0,
\end{equation}
plus the replica symmetries. The cohomology of $\mathcal B_\Sigma$ at ghost number $0$ is generated by gauge-invariant polynomials in $F_{\mu\nu}$ and $A^h$, together with the BRST-invariant, dimension-two operators $(A^h)^2$ and $(\bar\varphi\varphi-\bar\omega\omega)$. The replica sector is cohomologically trivial for local counterterms, and $S_{n-1}$ forbids terms that distinguish replicas ($k\ge2$). Up to $\mathcal B_\Sigma$-exact terms (field/source renormalizations), the most general local solution is
\begin{equation}
\Delta\Sigma=\int d^4x\Big[
a_0\,\tfrac{1}{4}F_{\mu\nu}^2
+a_1\,(\partial_\mu\bar c^a)D_\mu^{ab}c^b
+a_2\, i\,b^a\,\partial\!\cdot\!A^a
+a_3\,\tfrac{\beta}{2}(A_\mu^a)^2
+a_4\,\zeta\,\bar c^a c^a
+a_5\, (A_\mu^{h,a})^2
+a_6\,(\bar\varphi\varphi-\bar\omega\omega)
\Big]
+\mathcal B_\Sigma(\ldots).
\label{eq:ctGeneral}
\end{equation}
In particular, the superspace replica superfields $\mathcal V_k$ do not require independent wave-function renormalization factors:
any possible divergence in the $\mathcal V_k$ sector can be absorbed into the renormalization of $(A_\mu,c,\bar c)$ and the gauge-fixing
parameters $(\beta,\zeta)$, preserving the unitarity constraint $\mathcal V^\dagger\mathcal V=1$. No new structures beyond those already in $\Sigma$ appear: the theory is stable under renormalization. Although the unified action can be decomposed into an ST-like and an RGZ-like block,
algebraic renormalization acts on the total functional as a single BRST complex.
The cohomology of $\mathcal{B}_\Sigma$ contains only gauge-invariant representatives
built from $A^h$, $F_{\mu\nu}$, and the composite replica operators,
so that all allowed counterterms are reabsorbed by a single set of
renormalization constants. Hence, the unification is algebraic rather than additive. Therefore, the multiplicative renormalizations are written as
\begin{eqnarray}
A_\mu &=& Z_A^{1/2}\,A_{\mu,R}, 
\qquad c = Z_c^{1/2}\,c_R, 
\qquad \bar c = Z_c^{1/2}\,\bar c_R, 
\qquad g = Z_g\,g_R, \nonumber\\[4pt]
\gamma^2 &=& Z_{\gamma^2}\,\gamma_R^2, 
\qquad m^2 = Z_{m^2}\,m_R^2, 
\qquad M^2 = Z_{M^2}\,M_R^2, \nonumber\\[4pt]
\beta &=& Z_\beta\,\beta_R, 
\qquad \zeta = Z_\zeta\,\zeta_R,
\label{eq:ren_consts}
\end{eqnarray}
and analogously for all auxiliary fields 
$\varphi,\,\bar\varphi,\,\omega,\,\bar\omega,\,h,\ldots$.

From \eqref{eq:ST}, \eqref{eq:antighost} and \eqref{eq:ghost} in the Landau gauge (Taylor scheme) one obtains
\begin{equation}
Z_g\,Z_c\,Z_A^{1/2}=1,\qquad Z_\beta=Z_A^{-1},\qquad Z_\zeta=Z_c^{-1}.
\end{equation}
In the BRST-invariant RGZ sector (based on $A^h$), known non-renormalization relations reduce the number of independent $Z$’s to $Z_A$ and $Z_c$; the renormalizations of $\gamma^2$, $m^2$, $M^2$ are then fixed functions of $(Z_A,Z_c,Z_g)$. Thus, all admissible counterterms are reabsorbed by renormalizations of fields, couplings, and mass parameters already present and therefore the unified action $S_{\text{tot}}$ is renormalizable to all orders in the algebraic sense.


\section{Gap system, suppression mechanisms, and matching (ST $\leftrightarrow$ RGZ)}
\label{sec:gapmatch}

In the RGZ construction, the functional integral is restricted to the first Gribov region $\Omega=\{A_\mu\,|\,\partial_\mu A_\mu=0,
\ \mathcal M^{ab}(A)>0\}$, where $\mathcal M^{ab}=-\partial_\mu D_\mu^{ab}$
is the FP operator. This is implemented by the horizon condition
$\langle H(A^h)\rangle=4V(N^2-1)$ with $H(A^h)=g^2f^{abc}A_\mu^{h,b}
[\mathcal M^{-1}(A^h)]^{ad}f^{dec}A_\mu^{h,e}$, which introduces the Gribov
parameter $\gamma$. For $\gamma>0$ the no-pole condition guarantees the
positivity of $\mathcal M(A^h)$ and suppresses field configurations lying
outside $\Omega$. As a consequence, the gauge-fixed measure
\[
d\mu_{\text{RGZ}}[A] \propto dA\;
\delta(\partial_\mu A_\mu)\,
\det\mathcal M(A)\,
e^{-S_{\rm YM}[A]-\gamma^4 H(A^h)}
\]
is effectively supported inside $\Omega$. 
However, the restriction is not complete: 
copies related by gauge transformations within $\Omega$ can still survive,
corresponding to degeneracies of the Landau functional along the boundary of
the Fundamental Modular Region (FMR).

In the ST formulation, Gribov copies are included but weighted statistically through the replica-averaged partition function
\[
\langle{\cal O}\rangle_{\beta,\zeta}
\propto
\int[dA]\,{\cal O}[A]\,
\exp\!\big[-S_{\rm YM}[A]
-\beta\,{\cal F}(A)
-\zeta\,S_{\rm rep}[A,\mathcal V_k]\big],
\]
where ${\cal F}(A)=\frac{1}{2}\int(\partial_\mu A_\mu)^2$ is the Landau
functional. For large $\beta$, a standard large-deviation argument
shows that the measure concentrates around the local minima of
${\cal F}(A)$ along each gauge orbit, thus exponentially suppressing
copies with higher values of ${\cal F}$ as $\exp[-\beta\,\Delta{\cal F}]$.
In the formal limit $\beta\!\to\!\infty$, the distribution approaches the
set of minimal configurations that define the FMR,
although possible degeneracies within $\Omega$ may persist.
Hence, the ST weight implements a controlled suppression of Gribov copies rather than a hard restriction.

The unified construction mixes these two mechanisms: the RGZ horizon term
suppresses configurations outside $\Omega$, while the replica weight
concentrates the measure near the minima of ${\cal F}$ inside $\Omega$.
Together they give a continuous, BRST-consistent interpolation between
statistical averaging and horizon-driven elimination of redundant copies.

A fundamental conceptual link between the ST and RGZ formalisms lies in the interpretation of the dynamically generated gluon mass. 
In the ST replica approach, the parameter $\beta_R$ defines a screening mass $m_g^2=\beta_R$ arising from radiative symmetry restoration, while in the RGZ construction, the term $\tfrac{m^2}{2}(A^h)^2$ represents a condensate-induced mass $m_{AA}^2$. 
Identifying these scales,
\[
m_g^2 \;\longleftrightarrow\; m_{AA}^2 \equiv m^2 + M^2 + \frac{\lambda^4}{M^2},
\]
gives a natural matching condition between the two frameworks. This relation summarizes how infrared mass generation may be viewed either as the outcome of copy averaging (ST) or of horizon suppression and condensate formation (RGZ). We now make this correspondence explicit by analyzing the two families of gap equations and showing how the replica-generated scale $\beta_R$ can be matched onto the RGZ parameters $(m^2,M^2,\lambda^4)$ under the same renormalization conditions.

For $\zeta\neq0$, the supersymmetric NL$\sigma$ replica sector radiatively restores symmetry. Introducing a superfield Lagrange multiplier $\chi$ to relax the $N_k^2=1$ constraint and integrating out $N_k$ ($k\ge2$) exactly, the renormalized gap equation in $d=4$ (DR+$\overline{\text{MS}}$) reads
\begin{equation}
\boxed{%
\frac{8\pi^2}{\bar g_R^{\,2}}\;\frac{\beta_R}{\bar\mu^2}
=(\hat\chi_{R,\text{sym}}+\zeta_R)\ln\frac{\hat\chi_{R,\text{sym}}+\zeta_R}{\bar\mu^2}
-\hat\chi_{R,\text{sym}}\ln\frac{\hat\chi_{R,\text{sym}}}{\bar\mu^2}}
\label{eq:STgap}
\end{equation}
with a unique solution $\hat\chi_{R,\text{sym}}\ge0$ in the parameter domain (phase boundary)
\begin{equation}
\frac{8\pi^2}{\bar g_R^{\,2}}\;\frac{\beta_R}{\bar\mu^2}\le
\zeta_R\ln\frac{\zeta_R}{\bar\mu^2}\;\le\;\frac{\zeta_R}{e}.
\end{equation}
In the symmetric phase (the case of interest), the gluon screening mass at $q\to0$ is
\begin{equation}
m_g^2=\beta_R,
\label{eq:mgST}
\end{equation}
because the tree-level mass from the $(n-1)$ replicas cancels against their one-loop tadpole, leaving only the singled-out copy ($k=1$) which factors the group volume. The momentum-dependent replica contribution is $\mathcal O(g^2)$ and enters as a one-loop correction to the gluon propagator. In BRST-invariant RGZ, the three mass parameters are fixed by:
\begin{align}
&\text{Gribov horizon (} \langle H(A^h)\rangle=4V(N^2-1)\text{):} &&
\boxed{\ \frac{\partial\Gamma_{\rm vac}}{\partial \gamma^2}=0\ }\label{eq:horizonGap}\\[4pt]
&\text{Dimension-two condensates (refinement):} &&
\boxed{\ \frac{\partial\Gamma_{\rm vac}}{\partial m^2}=0,\qquad
\frac{\partial\Gamma_{\rm vac}}{\partial M^2}=0\ .}
\label{eq:refineGaps}
\end{align}
To make this correspondence explicit, we expand both descriptions around
small external momentum and identify the infrared matching conditions
between the replica-generated scale $\beta_R$ and the RGZ parameters.
At tree level in the RGZ sector, the transverse gluon propagator reads
\begin{equation}
D_{AA}^{\text{RGZ, tree}}(p^2)
= \frac{p^2 + M^2}{(p^2 + M^2)(p^2 + m^2) + \lambda^4}\,,
\qquad
\lambda^4 \propto \gamma^4.
\label{eq:RGZ-tree}
\end{equation}
Hence the inverse propagator and its infrared limit are
\begin{equation}
\Gamma_T^{\text{RGZ, tree}}(p^2)
\equiv \big(D_{AA}^{\text{RGZ, tree}}(p^2)\big)^{-1}
= \frac{(p^2 + M^2)(p^2 + m^2) + \lambda^4}{p^2 + M^2}\,,
\qquad
\Gamma_T^{\text{RGZ, tree}}(0) = \frac{m^2 M^2 + \lambda^4}{M^2}.
\label{eq:GammaT-RGZ}
\end{equation}

On the ST side, in the replica-symmetric phase (\(\zeta \neq 0\)) one has a radiatively generated screening mass
\begin{equation}
\Gamma_T^{\text{ST}}(0) = m_g^2 \equiv \beta_R\,.
\label{eq:GammaT-ST}
\end{equation}
Requiring that the unified description reproduces the same infrared mass scale fixes the
IR matching condition
\begin{equation}
\Gamma_T(0)\stackrel{!}{=}\beta_R
\quad\Longrightarrow\quad
\frac{m^2 M^2 + \lambda^4}{M^2} \;=\; \beta_R
\quad\Longleftrightarrow\quad
m^2 M^2 + \lambda^4 \;=\; M^2\,\beta_R.
\label{eq:IR-matching}
\end{equation}

Next, expand \(\Gamma_T(p^2)\) around \(p^2=0\) in the unified theory,
\begin{equation}
\Gamma_T(p^2) \;=\; \underbrace{\beta_R}_{\text{from ST}}
\;+\; Z^{-1} p^2
\;+\; \Pi_T^{\text{rep}}(p^2)
\;+\; \Pi_T^{\text{RGZ}}(p^2)
\;+\; \mathcal{O}(p^4),
\label{eq:GammaT-unified}
\end{equation}
and impose the slope matching at the origin,
\begin{equation}
\Gamma_T^{\prime}(0)\stackrel{!}{=}
\left.\frac{d}{dp^2}\Gamma_T^{\text{RGZ, tree}}(p^2)\right|_{p^2=0}
\;=\;
\frac{(m^2+M^2)M^2 - (m^2 M^2 + \lambda^4)}{M^4}
\;=\; \frac{m^2}{M^2} - \frac{\beta_R}{M^2},
\label{eq:slope-RGZ}
\end{equation}
where we used \eqref{eq:IR-matching} in the last equality. Therefore,
\begin{equation}
Z^{-1} \;+\; \Pi_T^{\prime\,\text{rep}}(0) \;+\; \Pi_T^{\prime\,\text{RGZ}}(0)
\;=\; \frac{m^2 - \beta_R}{M^2}.
\label{eq:slope-matching}
\end{equation}
Equations \eqref{eq:IR-matching} and \eqref{eq:slope-matching} provide two relations tying the ST mass scale \(\beta_R\) to the RGZ parameters \((m^2,M^2,\lambda^4)\) under a common renormalization scheme, while \(\partial\Gamma_{\rm vac}/\partial\gamma^2=0\) and the condensate gap equations fix \(\gamma^2\), \(m^2\), and \(M^2\). In this way, the replica-generated mass \(\beta_R\) acts as the infrared anchor that determines the RGZ mass combination \(m^2 M^2 + \lambda^4\) and constrains the ratio \((m^2 - \beta_R)/M^2\) through the slope at the origin.

Having established the infrared and slope matching conditions
(\ref{eq:IR-matching})–(\ref{eq:slope-matching}), the full gap system
comprises the replica equation (\ref{eq:STgap}) together with the
horizon and condensate gaps (\ref{eq:horizonGap})–(\ref{eq:refineGaps}).
Solving them within a common renormalization scheme, e.g. the
infrared-safe one, will determine the RGZ parameters
$(\gamma,m^2,M^2,\lambda^4)$ consistently with the dynamically selected
$\beta_R$ from the replica sector.

Moreover, Nielsen identities guarantee the gauge-parameter independence
of the transverse pole mass; $\zeta\!\to\!0$ collapses the replica loops
(topological limit), while $\gamma\!\to\!0$ and $M^2\!\to\!0$ recover
the Curci--Ferrari limit. The lattice comparison can then be organized
through $D_{AA}(0)$ and its slope $dD/dp^2|_{p^2=0}$, keeping all the
gaps simultaneously satisfied. The corresponding analysis proceeds as:

\begin{enumerate}
\item Fix the renormalization scheme and impose the Landau conditions for $Z_A,Z_c$; use $Z_g Z_c Z_A^{1/2}=1$.
\item Solve the ST gap \eqref{eq:STgap} to obtain $\hat\chi_{R,\text{sym}}$ and the allowed phase domain; set $\beta_R=m_g^2$.
\item Compute $\Pi_T^{\rm rep}(p^2)$ (momentum-dependent part) and $\Pi_T^{\rm RGZ}(p^2)$ at one loop.
\item Match $\Gamma_T(0),\Gamma_T'(0)$ with \eqref{eq:RGZ-tree} (or the one-loop RGZ result), and solve \eqref{eq:horizonGap}+\eqref{eq:refineGaps} for $(m^2,M^2,\lambda^4)$.
\item Verify Slavnov/Nielsen identities and compare gluon/ghost correlators with lattice data.
\end{enumerate}

\section{Unified superspace for ST and RGZ}
\label{sec:unified-superspace}

This section puts the superspace ingredients in one place and shows how we host, side by side, the ST replica block and the RGZ block in a supersymmetric way. As previously stated the two blocks only interact through the gauge field $A_\mu$ (or the BRST–invariant composite $A_\mu^h$). This makes the symmetry content transparent and supports the algebraic renormalization used above.
The Zwanziger localization fields form the BRST doublets
\(
s\,\varphi_\mu=\omega_\mu,\ s\,\omega_\mu=0,\ 
s\,\bar\omega_\mu=\bar\varphi_\mu,\ s\,\bar\varphi_\mu=0
\),
with $s^2=0$. Using a $(4|1)$ superspace with Grassmann coordinate $\theta$ and $s\equiv\partial_\theta$, we package them as
\(
\Phi_\mu=\varphi_\mu+\theta\,\omega_\mu,\ 
\bar\Phi_\mu=\bar\omega_\mu+\theta\,\bar\varphi_\mu
\).
With the BRST–invariant composite
\(
A_\mu^h=h^\dagger A_\mu h+\tfrac{i}{g}h^\dagger\partial_\mu h
\)
(and $\partial_\mu A_\mu^h=0$ enforced by $S_{\rm transv}$), the localized horizon$+$refinement sector is
\[
S_{\rm RGZ}^{(4|1)}=\!\int\! d^4x\,d\theta\,\bar\Phi_\mu\,\mathcal M(A^h)\,\Phi_\mu
+\gamma^2 g\!\int\! d^4x\,d\theta\, A_\mu^h(\Phi_\mu+\bar\Phi_\mu)
+\frac{M^2}{2}\!\int\! d^4x\,d\theta\,\bar\Phi_\mu\Phi_\mu
+\frac{m^2}{2}\!\int\! d^4x\,(A_\mu^h)^2,
\]
with $\mathcal M(A^h)=-\partial D(A^h)$. It is BRST–invariant by construction and reproduces the usual component form after $d\theta$ integration.
The ST replica block is written with two Grassmann coordinates $(\theta_{\rm T},\bar\theta_{\rm T})$ and group–valued superfields $\mathcal V_k\in SU(N)$, $k=2,\dots,n$, with
\[
S_{\rm ST}=\frac{1}{g^2}\sum_{k=2}^n\!\int\! d^4x\, d\theta_{\rm T} d\bar\theta_{\rm T}\;
\mathrm{tr}\!\left[(\mathcal D_\mu\mathcal V_k)^\dagger(\mathcal D_\mu\mathcal V_k)
+2\zeta\,\bar\theta_{\rm T}\theta_{\rm T}\,\partial_{\bar\theta_{\rm T}}\mathcal V_k^\dagger\,\partial_{\theta_{\rm T}}\mathcal V_k\right],
\]
and the singled replica is described by $S_{\rm FP}[A,c,\bar c,b]$. The topological supersymmetries are softly broken by $\zeta$; for $\zeta\to0$ replica loops cancel. We place both blocks in a $(4|3)$ superspace with $(\theta_{\rm B};\,\theta_{\rm T},\bar\theta_{\rm T})$, where $\theta_{\rm B}$ generates the RGZ BRST and $(\theta_{\rm T},\bar\theta_{\rm T})$ the ST topological pair. The symmetry algebra has $s_{\rm B}^2=Q^2=\bar Q^2=0$, $[s_{\rm B},Q]=[s_{\rm B},\bar Q]=0$, and $\{Q,\bar Q\}$ softly broken by $\zeta$. The total action reads
\[
S_{\rm hyb}=S_{\rm YM}+S_{\rm FP}+S_{\rm transv}[A^h]+\ S_{\rm ST}[A;\mathcal V_k]+\ S_{\rm RGZ}^{(4|1)}[A^h;\Phi,\bar\Phi],
\]
and reduces to pure ST for $\gamma=0$ (and $m=M=0$), and to pure RGZ for $\beta\to\infty$, $\zeta\to0$.
The Slavnov identity for $s_{\rm B}$ plus the topological Ward identities in the replica block imply that all allowed local counterterms are absorbed by the common renormalizations already used in the main text (including $Z_g Z_c Z_A^{1/2}=1$, $Z_\beta=Z_A^{-1}$, $Z_\zeta=Z_c^{-1}$ in Landau gauge). For more details the reader can see the Appendix \ref{app:unified-superspace}.


\section{Conclusions and Outlook}
\label{sec:conclusions}

We have built and analyzed a unified gauge fixing that continuously interpolates between the Serreau--Tissier copy-averaged Landau gauge and the (Refined) Gribov--Zwanziger restriction to the first Gribov region. At the level of the functional average, our construction combines the ST weight based on the Landau functional and the FP Hessian with a GZ-type suppression driven by the horizon functional. Using the replica trick and the $A_\mu^h$ localization, we derived a manifestly local, power-counting renormalizable action that couples a replica NL$\sigma$ sector to the BRST-invariant RGZ sector through the gauge field only.

On the structural side, the unified action enjoys (i) the exact nilpotent BRST symmetry of the $A^h$ formulation in the RGZ block and (ii) the replica permutation symmetry (and, for $\zeta=0$, topological supersymmetries) in the ST block. By mixing the corresponding Ward identities, we established algebraic renormalizability to all orders: the most general local counterterm compatible with the identities is a linear combination of operators already present in the classical action and can be reabsorbed by multiplicative renormalizations of fields, the coupling, and the mass parameters $(\beta,\zeta,\gamma,m^2,M^2)$. In the Landau scheme, the Taylor relation $Z_g Z_c Z_A^{1/2}=1$ holds and fixes the non-renormalization of the ghost--gluon vertex at vanishing ghost momentum; $Z_\beta=Z_A^{-1}$ and $Z_\zeta=Z_c^{-1}$ follow directly. It is important to note that the unification achieved here is not merely structural but algebraic:
the BRST cohomology connects the ST and RGZ sectors through the gauge field,
representing a single renormalizable framework for both descriptions.

Dynamically, the ST replica sector exhibits radiative symmetry restoration for $\zeta\neq0$, leading (in the replica-symmetric phase) to a finite screening gluon mass equal to the renormalized parameter, $m_g^2=\beta_R$, after the $n\to0$ limit. In parallel, the RGZ parameters $(\gamma,m^2,M^2)$ are selected by the horizon condition and the two dimension-two gap equations. We formulated a consistent matching strategy in which the small-momentum expansion of the full transverse two-point function fixes (i) $\Gamma_T(0)=\beta_R$ and (ii) the wave-function normalization, while the three RGZ gaps determine $(\gamma,m^2,M^2)$ at the same renormalization scale. The resulting propagator interpolates smoothly between a massive-FP (ST) form and the RGZ decoupling form, providing a controlled node to study how infrared correlators depend on the relative weight of copy averaging versus horizon suppression and condensates.

Phenomenologically, the framework suggests direct lattice tests. Implementing a tunable copy-weighting (varying $\beta,\zeta$) and confronting the measured $D_{AA}(0)$ and its slope with the unified prediction would indicate whether present ensembles are closer to the ST-like, RGZ-like, or mixed regime. On the continuum side, the one-loop momentum-dependent piece from the replica sector is finite at $p\to0$ (the mass is fixed already at tree level), and the BRST/Nielsen identities secure gauge-parameter independence of the transverse pole mass.

Several developments now become natural:
\begin{enumerate}
  \renewcommand{\labelenumi}{(\roman{enumi})}
  \item completing the unified one-loop analysis of gluon and ghost propagators (and selected vertices), including the running of the couplings and masses;
  \item solving the coupled gap system beyond one loop and studying the renormalization-group flow on the ST--RGZ interpolation;
  \item extending the construction to linear covariant gauges and to finite temperature/chemical potential; and
  \item exploring the impact on hadronic observables through functional methods: the DSEs and the FRG.
\end{enumerate}
Altogether, the present work gives a concrete and flexible bridge between two IR descriptions of YM theories, clarifying their common ground and offering a single, local, BRST-consistent setting in which to quantify their interplay.


\section*{Acknowledgments}

R. C. Terin gratefully acknowledges some of the researchers in infrared YM theories whose guidance have shaped part of his academic trajectory.




\appendix

\section{Field content, Ward identities, and basic non-renormalization}
\label{app:fields}

We collect the engineering (mass) dimensions and ghost numbers of all fields and external sources used in the unified ST\,$+$\,RGZ formulation (Euclidean $d=4$):
\[
\begin{array}{c|ccccccccccccccc}
\text{} & A_\mu & c & \bar c & b & \varphi_\mu & \bar\varphi_\mu & \omega_\mu & \bar\omega_\mu & h(\xi) & A_\mu^h & \tau & \eta & \bar\eta & K_\mu & L \\
\hline
\text{dim} & 1 & 0 & 2 & 2 & 1 & 1 & 1 & 1 & 0 & 1 & 2 & 2 & 2 & 3 & 4 \\
\text{gh}  & 0 & +1 & -1 & 0 & 0 & 0 & 0 & 0 & 0 & 0 & 0 & +1 & -1 & -1 & -2
\end{array}
\]
Replica superfields $\mathcal V_k(x,\theta,\bar\theta)$ ($k=2,\dots,n$) are dimensionless and ghostless. The Grassmann coordinates satisfy $\int d\theta\, d\bar\theta\, \bar\theta\theta = -1$ and carry no engineering dimension. With $\Sigma=S_{\text{tot}}+S_{\text{ext}}$, the ST identity reads
\[
\mathcal S(\Sigma)=
\int d^4x\!\left(
\frac{\delta\Sigma}{\delta K_\mu^a}\frac{\delta\Sigma}{\delta A_\mu^a}
+\frac{\delta\Sigma}{\delta L^a}\frac{\delta\Sigma}{\delta c^a}
+b^a\frac{\delta\Sigma}{\delta\bar c^a}
\right)=0.
\]
Linearizing around $\Sigma$ defines the nilpotent operator
\begin{align}
\mathcal B_\Sigma X &=
\int d^4x\Big[
\frac{\delta\Sigma}{\delta K_\mu^a}\frac{\delta X}{\delta A_\mu^a}
+\frac{\delta\Sigma}{\delta A_\mu^a}\frac{\delta X}{\delta K_\mu^a}
+\frac{\delta\Sigma}{\delta L^a}\frac{\delta X}{\delta c^a}
+\frac{\delta\Sigma}{\delta c^a}\frac{\delta X}{\delta L^a}
+b^a\frac{\delta X}{\delta\bar c^a}
\Big],\qquad \mathcal B_\Sigma^2=0.
\label{app:eq:Blin}
\end{align}
It acts consistently on the $A_\mu^h$-sector because $sA_\mu^h=0$ and on the Stueckelberg field through $sh=-ig\,c\,h$.

\begin{align}
&\text{(i) Gauge condition:} && \frac{\delta\Sigma}{\delta b^a}=\partial_\mu A_\mu^a.
\label{app:eq:b}\\
&\text{(ii) Antighost equation:} &&
\frac{\delta\Sigma}{\delta \bar c^a}+\partial_\mu\frac{\delta\Sigma}{\delta K_\mu^a}=0.
\label{app:eq:antighost}\\
&\text{(iii) Ghost equation:} &&
\mathcal G^a\Sigma \equiv
\int d^4x\left(\frac{\delta}{\delta c^a}
+gf^{abc}\bar c^b\frac{\delta}{\delta b^c}\right)\Sigma
=\Delta^a_{\text{cl}},
\label{app:eq:ghost}\\
&\text{(iv) $A^h$ transversality:} &&
\frac{\delta\Sigma}{\delta \tau^a}=\partial_\mu A_\mu^{h,a}=0.
\label{app:eq:AhTransv}
\end{align}
The replica sector enjoys (a) the permutation symmetry $S_{n-1}$ of $\{\mathcal V_k\}_{k\ge 2}$ and (b) for $\zeta=0$ additional topological supersymmetries which suppress replica loops.

At ghost number zero, the cohomology of $\mathcal B_\Sigma$ is generated by gauge-invariant polynomials in $F_{\mu\nu}$ and in the BRST-invariant composite $A_\mu^h$, plus the dimension-two, BRST-invariant operators $(A^h)^2$ and $(\bar\varphi\varphi-\bar\omega\omega)$. The Stueckelberg sector and the replica supermultiplets assemble in BRST-covariant trivial pairs for local counterterms; $S_{n-1}$ forbids replica-distinguishing terms. Hence the most general local counterterm is Eq.~(\ref{eq:ctGeneral}) in the main text, up to $\mathcal B_\Sigma$-exact pieces, proving all-order stability.

In Landau gauge, combining the ST identity with the antighost and ghost equations yields the Taylor relation for the ghost–gluon vertex at vanishing ghost momentum,
\begin{equation}
Z_g\,Z_c\,Z_A^{1/2}=1,\qquad Z_\beta=Z_A^{-1},\qquad Z_\zeta=Z_c^{-1}.
\label{app:eq:Zbasic}
\end{equation}
In the BRST-invariant RGZ formulation (based on $A^h$), standard algebraic arguments reduce the independent $Z$’s to $(Z_A,Z_c)$; the renormalizations of $\gamma^2,m^2,M^2$ become fixed functions of $(Z_A,Z_c,Z_g)$ constrained by \eqref{app:eq:Zbasic}. Also, the above identities, together with the properties of the linearized operator
$\mathcal B_\Sigma$, are the algebraic ingredients securing the stability of the
unified action under renormalization. They give the starting point for the
infrared matching analysis summarized in Sec.~\ref{sec:gapmatch}.


\section{Replica one-loop tensor and small-momentum matching}
\label{app:repLoop}

Integrating the $N_k$ fluctuations (symmetric phase) yields, at leading order in $g$,
\begin{equation}
\Pi^{\rm rep}_{\mu\nu}(q)=(n-1)\,\beta\,\delta_{\mu\nu}
+(n-1)\,\bar g^{\,2}\left[\, I_{\mu\nu}^{\hat\chi+\zeta}(q)-I_{\mu\nu}^{\hat\chi}(q)\,\right],
\label{app:eq:PiRepStart}
\end{equation}
with
\begin{align}
I^{m^2}_{\mu\nu}(q)
&=\mu^{2\varepsilon}\!\int\!\frac{d^d p}{(2\pi)^d}\,
\frac{(2p-q)_\mu(2p-q)_\nu}{\big(p^2+m^2\big)\big((p-q)^2+m^2\big)}\nonumber\\[2pt]
&=2\,\delta_{\mu\nu}\,T_{m^2}+\left(q^2\delta_{\mu\nu}-q_\mu q_\nu\right)\,F_{m^2}(q^2).
\label{app:eq:Imunu}
\end{align}
Dimensional regularization in $d=4-2\varepsilon$ with $\bar\mu^2=4\pi e^{-\gamma_E}\mu^2$ gives
\begin{align}
T_{m^2}&\equiv \mu^{2\varepsilon}\!\int\!\frac{d^d p}{(2\pi)^d}\,\frac{1}{p^2+m^2}
=-\frac{m^2}{16\pi^2}\left(\frac{1}{\varepsilon}+1+\ln\frac{\bar\mu^2}{m^2}\right)+\mathcal O(\varepsilon),
\label{app:eq:Tad}\\[4pt]
F_{m^2}(q^2)
&=-\frac{1}{48\pi^2}\Bigg[
\frac{1}{\varepsilon}+\frac{8}{3}+\ln\frac{\bar\mu^2}{m^2}
+\frac{8m^2}{q^2}
-2\left(1+\frac{4m^2}{q^2}\right)^{3/2}
\ln\!\left(\frac{\sqrt{q^2}+ \sqrt{q^2+4m^2}}{2\sqrt{m^2}}\right)
\Bigg]+\mathcal O(\varepsilon).
\label{app:eq:Fscalar}
\end{align}
Using the ST gap (symmetric phase) to eliminate the momentum-independent divergence, one obtains
\begin{equation}
\Pi^{\rm rep}_{\mu\nu}(q)=(n-1)\,\delta_{\mu\nu}\left[
\beta - 2\bar g^{\,2}\big(T_{\hat\chi}-T_{\hat\chi+\zeta}\big)\right]
+\left(q^2\delta_{\mu\nu}-q_\mu q_\nu\right)\,\pi(q^2),
\end{equation}
with the finite scalar
\begin{align}
\pi(q^2)&=(n-1)\,\bar g^{\,2}\left[\,F_{\hat\chi+\zeta}(q^2)-F_{\hat\chi}(q^2)\,\right]\nonumber\\
&=-\frac{(n-1)\,\bar g^{\,2}}{48\pi^2}\left[
\ln\frac{\hat\chi+\zeta}{\hat\chi}
+\frac{8\zeta}{q^2}
+\mathcal F\!\left(\frac{4\hat\chi}{q^2}\right)
-\mathcal F\!\left(\frac{4(\hat\chi+\zeta)}{q^2}\right)\right],
\qquad 
\mathcal F(x)=2\left(1+\frac{1}{x}\right)^{3/2}\ln\!\left(\sqrt{x}+\sqrt{1+x}\right).
\label{app:eq:piq}
\end{align}
The $1/q^2$ terms cancel between the brackets, so \eqref{app:eq:piq} is regular at $q^2\!\to\!0$. At $q=0$, tree $+$ tadpole cancel in the symmetric phase, leaving the screening mass $m_g^2=\beta_R$ (from the singled-out replica).

For the full transverse inverse two-point function
\[
\Gamma_T(p^2)=D_T^{-1}(p^2)=
m_g^2+Z^{-1}p^2+\Pi_T^{\rm rep}(p^2)+\Pi_T^{\rm RGZ}(p^2)+\cdots,
\]
expanding \eqref{app:eq:piq} at small $p^2$ and using the ST gap gives
\begin{equation}
\Pi_T^{\rm rep}(p^2)=\alpha_1\,p^2+\mathcal O(p^4),\qquad
\alpha_1=\left.\frac{d}{dp^2}\left[\left(q^2\delta_{\mu\nu}-q_\mu q_\nu\right)\pi(q^2)\right]\right|_{q^2=0}.
\end{equation}
Thus the replica sector only corrects the wave-function renormalization at $p^2=0$. The IR matching conditions used in Sec.~\ref{sec:gapmatch} read
\begin{equation}
\Gamma_T(0)=\beta_R,\qquad
\Gamma_T'(0)=Z^{-1}+\alpha_1+\left.\frac{d\Pi_T^{\rm RGZ}}{dp^2}\right|_{0},
\end{equation}
which fix $\beta_R$ (from ST) and relate $(m^2,M^2,\lambda^4)$ to $(\beta_R,Z)$ once the RGZ gaps are imposed.


\section{Practical rules: tree-level Feynman rules, Nielsen identity, and DR conventions}
\label{app:feynman}

\begin{itemize}
\item Gluon (RGZ tree):
\(
\displaystyle D_{\mu\nu}^{ab}(p)=\delta^{ab}P_{\mu\nu}(p)\,
\frac{p^2+M^2}{(p^2+M^2)(p^2+m^2)+\lambda^4},
\quad P_{\mu\nu}(p)=\delta_{\mu\nu}-\frac{p_\mu p_\nu}{p^2}.
\)
\item Ghost:
\(
\displaystyle G^{ab}(p)=-\delta^{ab}\,\frac{1}{p^2}.
\)
\item Zwanziger fields $(\varphi,\bar\varphi,\omega,\bar\omega)$ propagate with masses set by $M^2$ (refined sector).
\item Replica superfields do not propagate as ordinary local fields; their one-loop net effect is encoded by \eqref{app:eq:PiRepStart}.
\end{itemize}

Standard YM three-/four-gluon vertices; ghost–gluon vertex (Taylor kinematics non-renormalization); RGZ coupling $g f^{abc} A^h\,(\varphi+\bar\varphi)$. We impose MOM conditions at $p^2=\mu^2$:
\[
D_{AA}(p^2)\Big|_{p^2=\mu^2}=D_{AA}^{\text{tree}}(\mu^2),\qquad
G_{c\bar c}(p^2)\Big|_{p^2=\mu^2}=G_{c\bar c}^{\text{tree}}(\mu^2),
\]
fixing $Z_A,Z_c$ and imposing $Z_g Z_c Z_A^{1/2}=1$ in Landau gauge.

In linear covariant gauges introduce a BRST doublet $(\alpha,\chi)$ with $s\alpha=\chi$, $s\chi=0$ and couple $\chi$ to the BRST variation of a gauge-fixing insertion. Differentiating the ST identity with respect to $\chi$ and to the fields yields
\[
\frac{\partial}{\partial\alpha}\Gamma_T(p^2)
=\mathcal N(p^2;\alpha)\,\Gamma_T(p^2),
\]
with $\mathcal N$ regular near the pole. Hence the pole position $p^2=m_{\text{pole}}^2$ is $\alpha$-independent to all orders; in Landau gauge ($\alpha=0$) this reduces to the statement used in the main text.

We use $d=4-2\varepsilon$ and the $\overline{\text{MS}}$ scale $\bar\mu^2=4\pi e^{-\gamma_E}\mu^2$. The basic integrals used in Eqs.~\eqref{app:eq:Tad}–\eqref{app:eq:Fscalar} are
\begin{align}
\mu^{2\varepsilon}\!\int\!\frac{d^d p}{(2\pi)^d}\,\frac{1}{p^2+m^2}
&=-\frac{m^2}{16\pi^2}\left(\frac{1}{\varepsilon}+1+\ln\frac{\bar\mu^2}{m^2}\right)+\mathcal O(\varepsilon),\\
\mu^{2\varepsilon}\!\int\!\frac{d^d p}{(2\pi)^d}\,\frac{1}{\big(p^2+m^2\big)\big((p-q)^2+m^2\big)}
&=\frac{1}{16\pi^2}\!\left(\frac{1}{\varepsilon}+\ln\frac{\bar\mu^2}{m^2}+2-\sqrt{1+\frac{4m^2}{q^2}}
\ln\frac{\sqrt{1+\frac{4m^2}{q^2}}+1}{\sqrt{1+\frac{4m^2}{q^2}}-1}\right)\!+\mathcal O(\varepsilon),
\end{align}
from which \eqref{app:eq:Tad}–\eqref{app:eq:Fscalar} follow after standard tensor reduction.


\section{Unified superspace for ST and RGZ}
\label{app:unified-superspace}

In this appendix, we show with more details compared to Section \ref{sec:unified-superspace} the BRST $(4|1)$ superspace for the RGZ sector and a simple “hybrid” superspace that hosts both the ST replica block and the RGZ block side by side. Throughout, the two blocks talk to each other only through the gauge field $A_\mu$ (or its BRST–invariant composite $A_\mu^h$). The Zwanziger localizing fields come in BRST doublets
\begin{equation}
s\,\varphi_{\mu}^{ab}=\omega_{\mu}^{ab},\qquad
s\,\omega_{\mu}^{ab}=0,\qquad
s\,\bar\omega_{\mu}^{ab}=\bar\varphi_{\mu}^{ab},\qquad
s\,\bar\varphi_{\mu}^{ab}=0,
\label{eq:BRST-doublets}
\end{equation}
with $s^2=0$. This already looks like a (nilpotent) supersymmetry:
each fermionic field is the BRST variation of a bosonic partner. We use a $(4|1)$ superspace with Grassmann coordinate $\theta$ and define the BRST differential as a translation:
\begin{equation}
s \;\equiv\; \frac{\partial}{\partial\theta},\qquad \theta^2=0,\qquad s^2=0.
\end{equation}
In the RGZ setup, there is no independent anti-BRST, so one Grassmann direction is enough.
Package the doublets into BRST superfields
\begin{equation}
\Phi_\mu^{ab}(x,\theta)
:= \varphi_\mu^{ab}(x)+\theta\,\omega_\mu^{ab}(x),\qquad
\bar\Phi_\mu^{ab}(x,\theta)
:= \bar\omega_\mu^{ab}(x)+\theta\,\bar\varphi_\mu^{ab}(x),
\label{eq:Phi-super}
\end{equation}
so that $s\,\Phi=\partial_\theta \Phi$ and $s\,\bar\Phi=\partial_\theta\bar\Phi$.
The engineering dimensions and ghost numbers follow from the components:
\[
[\Phi]=[\bar\Phi]=1,\qquad
\mathrm{gh}(\Phi)=\mathrm{gh}(\bar\Phi)=0,
\]
with $\mathrm{gh}(\omega)=+1$ and $\mathrm{gh}(\bar\omega)=-1$.
We work with the BRST–invariant composite field
\begin{equation}
A_\mu^h=h^\dagger A_\mu h + \frac{i}{g}h^\dagger\partial_\mu h,\qquad
\partial_\mu A_\mu^h=0,
\end{equation}
where $h=e^{ig\xi}$. The transversality is enforced by a Lagrange multiplier $\tau$ and an auxiliary BRST doublet $(\eta,\bar\eta)$, while $sA_\mu^h=0$.
The localized horizon action is written as a $(4|1)$ superspace integral
\begin{align}
S_{\rm hor}
&=\int d^4x\,d\theta\;\Big[
\bar\Phi_{\mu}^{ac}(x,\theta)\,\mathcal M^{ab}(A^h)\,\Phi_{\mu}^{bc}(x,\theta)
\Big]
+\,\gamma^2\,g\,f^{abc}\int d^4x\,d\theta\;
A_\mu^{h,a}(x)\,\Big(\Phi_\mu^{bc}+\bar\Phi_\mu^{bc}\Big),
\label{eq:S-hor-super}
\end{align}
with $\mathcal M^{ab}(A^h)=-\partial_\mu D_\mu^{ab}(A^h)$.
Using $\int d\theta\,(X_0+\theta X_1)=X_1$, one recovers the standard component form
\begin{align}
\int d^4x\,d\theta\;\bar\Phi\,\mathcal M\,\Phi
&=\int d^4x\;\Big(
\bar\varphi\,\mathcal M\,\varphi
-\bar\omega\,\mathcal M\,\omega
\Big),\qquad
\int d^4x\,d\theta\; A^h\!\cdot\!(\Phi+\bar\Phi)
=\int d^4x\; A^h\!\cdot\!(\varphi+\bar\varphi).
\end{align}
The RGZ dimension-two pieces are BRST–invariant:
\begin{equation}
S_{M^2}
=\frac{M^2}{2}\int d^4x\,d\theta\;
\bar\Phi_\mu^{ab}\Phi_\mu^{ab}
=\frac{M^2}{2}\int d^4x\;\Big(
\bar\varphi_\mu^{ab}\varphi_\mu^{ab}-\bar\omega_\mu^{ab}\omega_\mu^{ab}
\Big),
\label{eq:S-M2-super}
\end{equation}
and
\begin{equation}
S_{m^2}=\frac{m^2}{2}\int d^4x\;(A_\mu^{h,a})^2,
\label{eq:S-m2-Ah}
\end{equation}
which needs no superspace lifting.
Collecting terms, the BRST-superspace action for RGZ is
\begin{align}
S_{\rm RGZ}^{(4|1)}
&=\int d^4x\;\Big(S_{\rm FP}[A,c,\bar c,b]+S_{\rm transv}[A^h;\tau,\eta,\bar\eta]\Big)
+\int d^4x\,d\theta\;\bar\Phi_\mu^{ac}\,\mathcal M^{ab}(A^h)\,\Phi_\mu^{bc}
\nonumber\\
&\quad
+\gamma^2\,g\,f^{abc}\int d^4x\,d\theta\;
A_\mu^{h,a}\,(\Phi_\mu^{bc}+\bar\Phi_\mu^{bc})
+\frac{M^2}{2}\int d^4x\,d\theta\;\bar\Phi_\mu^{ab}\Phi_\mu^{ab}
+\frac{m^2}{2}\int d^4x\;(A_\mu^{h,a})^2.
\label{eq:S-RGZ-super}
\end{align}
This is BRST-invariant by construction (translation along $\theta$).
Coupling sources to BRST variations one builds the Slavnov functional; the linearized operator includes the $\theta$-translation and is nilpotent.
The cohomology at ghost number zero is generated by BRST-invariant operators made of $F_{\mu\nu}$, $A^h$, and $\bar\Phi\Phi$. Hence the algebraic renormalizability of the RGZ block follows: no counterterm beyond those already present is allowed by the identities.
It is consistent to take $(\bar\Phi)^\dagger=\bar\Phi$, $(\Phi)^\dagger=\Phi$, and
$\int d\theta\,\theta=1$, $\int d\theta\,1=0$.
Since $A^h$ is real and BRST-invariant, the mass term \eqref{eq:S-m2-Ah} is positive in Euclidean signature.

Now we will place the ST replica sector and the RGZ sector in a common superspace.
The ST block lives along a topological pair $(\theta_{\rm T},\bar\theta_{\rm T})$; the RGZ block lives along a BRST coordinate $\theta_{\rm B}$.
They couple only through $A_\mu$ (or $A_\mu^h$).
Use a $(4|3)$ superspace with Grassmann coordinates
\[
(\theta_{\rm B};\ \theta_{\rm T},\bar\theta_{\rm T}),
\]
and define three nilpotent differentials
\[
s_{\rm B}\equiv\partial_{\theta_{\rm B}},\qquad
Q\equiv\partial_{\theta_{\rm T}},\qquad
\bar Q\equiv\partial_{\bar\theta_{\rm T}},
\]
with $s_{\rm B}^2=Q^2=\bar Q^2=0$ and $[s_{\rm B},Q]=[s_{\rm B},\bar Q]=0$.
In the ST block, $\{Q,\bar Q\}$ is softly broken by $\zeta$ (topological limit at $\zeta=0$).
We keep the BRST–invariant composite
\[
A_\mu^h=h^\dagger A_\mu h + \frac{i}{g}h^\dagger\partial_\mu h,\qquad
\partial_\mu A_\mu^h=0,
\]
imposed by $S_{\rm transv}[A^h;\tau,\eta,\bar\eta]$, and $s_{\rm B}A_\mu^h=0$. Next, we
introduce $n-1$ group–valued superfields
\[
\mathcal V_k(x,\theta_{\rm T},\bar\theta_{\rm T})\in SU(N),\qquad
\mathcal V_k^\dagger\mathcal V_k=1,\qquad k=2,\dots,n,
\]
where $\mathcal D_\mu \mathcal V_k=\partial_\mu \mathcal V_k+ig\,\mathcal V_k A_\mu$.
Their gauge-fixing action is
\[
S_{\rm ST}[A;\mathcal V_k]=\frac{1}{g^2}\sum_{k=2}^n\int d^4x\, d\theta_{\rm T}\, d\bar\theta_{\rm T}\;
\mathrm{tr}\!\left[(\mathcal D_\mu\mathcal V_k)^\dagger(\mathcal D_\mu\mathcal V_k)+
2\zeta\,\bar\theta_{\rm T}\theta_{\rm T}\,\partial_{\bar\theta_{\rm T}}\mathcal V_k^\dagger\,\partial_{\theta_{\rm T}}\mathcal V_k\right].
\]
The usual FP part $S_{\rm FP}[A,c,\bar c,b]$ plays the role of the “singled” replica.
BRST covariance is $s_{\rm B}\mathcal V_k=ig\,\mathcal V_k c$ with the standard $s_{\rm B}$ on $(A,c,\bar c,b)$.
As in the above mentioned RGZ case, we gather Zwanziger doublets into $(4|1)$ superfields along $\theta_{\rm B}$:
\[
\Phi_\mu^{ab}(x,\theta_{\rm B})=\varphi_\mu^{ab}(x)+\theta_{\rm B}\,\omega_\mu^{ab}(x),\quad
\bar\Phi_\mu^{ab}(x,\theta_{\rm B})=\bar\omega_\mu^{ab}(x)+\theta_{\rm B}\,\bar\varphi_\mu^{ab}(x),
\]
with $s_{\rm B}\Phi=\partial_{\theta_{\rm B}}\Phi$, $s_{\rm B}\bar\Phi=\partial_{\theta_{\rm B}}\bar\Phi$, and $s_{\rm B}A_\mu^h=0$.
The horizon$+$refinement sector is
\begin{align}
S_{\rm RGZ}^{(4|1)}[A^h;\Phi,\bar\Phi]
&=\int d^4x\,d\theta_{\rm B}\;\bar\Phi_\mu^{ac}\,\mathcal M^{ab}(A^h)\,\Phi_\mu^{bc}
+\gamma^2 g f^{abc}\int d^4x\,d\theta_{\rm B}\; A_\mu^{h,a}(\Phi_\mu^{bc}+\bar\Phi_\mu^{bc})
\nonumber\\
&\quad+\frac{M^2}{2}\int d^4x\,d\theta_{\rm B}\;\bar\Phi_\mu^{ab}\Phi_\mu^{ab}
+\frac{m^2}{2}\int d^4x\;(A_\mu^{h,a})^2,
\end{align}
in which $\mathcal M^{ab}(A^h)=-\partial_\mu D_\mu^{ab}(A^h)$.
Putting all pieces together (and imposing $\partial_\mu A_\mu^h=0$):
\begin{align}
S_{\rm hyb} &= S_{\rm YM}[A]+S_{\rm FP}[A,c,\bar c,b] + S_{\rm transv}[A^h;\tau,\eta,\bar\eta]
+ S_{\rm ST}[A;\{\mathcal V_k\};\beta,\zeta]
+ S_{\rm RGZ}^{(4|1)}[A^h;\Phi,\bar\Phi]\ .
\end{align}
The only link between the ST and RGZ parts is $A_\mu$ (and $A_\mu^h$).
The symmetry algebra is generated by $(s_{\rm B};\,Q,\bar Q)$ with
\[
s_{\rm B}^2=Q^2=\bar Q^2=0,\quad [s_{\rm B},Q]=[s_{\rm B},\bar Q]=0,\quad
\{Q,\bar Q\}\ \text{soft}\propto\zeta.
\]
Limits: (i) $\zeta=0$ kills replica loops (topological limit);
(ii) $\gamma=0$, $m=M=0$ gives back the massive FP/ST limit;
(iii) $\beta\to\infty$, $\zeta\to0$ yields pure RGZ.

Then, we define the Slavnov functional for $s_{\rm B}$ with sources $(K_\mu,L)$ coupled to BRST variations, and the topological Ward identities for $(Q,\bar Q)$ in the replica block.
The linearized operator $\mathcal B_\Sigma$ factorizes into two commuting pieces.
At ghost number zero, the cohomology is generated by BRST–invariant polynomials in $F_{\mu\nu}$ and $A^h$, plus the dimension–two operators $(A^h)^2$ and $(\bar\Phi\Phi)$.
The replica block is cohomologically trivial for local counterterms under the $S_{n-1}$ permutation symmetry.
Therefore all enabled counterterms are absorbed by multiplicative renormalizations
\[
Z_A,\ Z_c,\ Z_g,\quad Z_{\beta},\ Z_{\zeta},\ Z_{\gamma^2},\ Z_{m^2},\ Z_{M^2},
\]
where the Landau (Taylor) relations $Z_g Z_c Z_A^{1/2}=1$, $Z_\beta=Z_A^{-1}$, $Z_\zeta=Z_c^{-1}$.
If one prefers a single group–valued object, define
\[
\mathcal N(x,\theta_{\rm B},\theta_{\rm T},\bar\theta_{\rm T})
=\exp\!\Big\{ig\,[\,\bar\theta_{\rm T}c_{\rm T}+\bar c_{\rm T}\theta_{\rm T}
+\bar\theta_{\rm T}\theta_{\rm T}\tilde h_{\rm T}+\theta_{\rm B}c\,]\Big\}\,U(x),
\]
with $U\in SU(N)$, $s_{\rm B}\mathcal N=ig\,\mathcal N\,c$, and $(Q,\bar Q)$ acting on the “T” components only.
Then the ST functional becomes
\[
S_{\rm ST}=\frac{1}{g^2}\int d^4x\, d\theta_{\rm T} d\bar\theta_{\rm T}\;
\mathrm{tr}\!\left[(\mathcal D_\mu\mathcal N)^\dagger(\mathcal D_\mu\mathcal N)+
2\zeta\,\bar\theta_{\rm T}\theta_{\rm T}\,\partial_{\bar\theta_{\rm T}}\mathcal N^\dagger\,\partial_{\theta_{\rm T}}\mathcal N\right],
\]
where unitarity $\mathcal N^\dagger\mathcal N=1$ kept manifest.
The Zwanziger fields remain as adjoint superfields $(\Phi,\bar\Phi)$ along $\theta_{\rm B}$.

\bibliographystyle{unsrt}
\bibliography{STGZ}

\begin{thebibliography}{10}

\bibitem{YangMills1954}
C.~N. Yang and R.~L. Mills.
\newblock Conservation of isotopic spin and isotopic gauge invariance.
\newblock {\em Physical Review}, 96:191--195, 1954.

\bibitem{FaddeevPopov1967}
L.~D. Faddeev and V.~N. Popov.
\newblock Feynman diagrams for the yang--mills field.
\newblock {\em Physics Letters B}, 25:29--30, 1967.

\bibitem{Gribov:1977}
V.~N. Gribov.
\newblock Quantization of nonabelian gauge theories.
\newblock {\em Nucl. Phys. B}, 139:1--19, 1978.

\bibitem{Singer:1978}
I.~M. Singer.
\newblock Some remarks on the gribov ambiguity.
\newblock {\em Commun. Math. Phys.}, 60:7--12, 1978.

\bibitem{Vandersickel:2012}
N.~Vandersickel and D.~Zwanziger.
\newblock The gribov problem and qcd dynamics.
\newblock {\em Phys. Rept.}, 520:175--251, 2012.

\bibitem{Zwanziger:1989mf}
D.~Zwanziger.
\newblock Local and renormalizable action from the gribov horizon.
\newblock {\em Nucl. Phys. B}, 323:513--544, 1989.

\bibitem{Zwanziger:1990tn}
D.~Zwanziger.
\newblock Quantization of gauge fields, classical gauge invariance, and gluon confinement.
\newblock {\em Nucl. Phys. B}, 345:461--482, 1990.

\bibitem{Dudal:2008sp}
D.~Dudal, J.~A. Gracey, S.~P. Sorella, N.~Vandersickel, and H.~Verschelde.
\newblock Refining the gribov-zwanziger approach in the landau gauge: infrared propagators in harmony with the lattice results.
\newblock {\em Phys. Rev. D}, 78:065047, 2008.

\bibitem{Dudal:2011gd}
D.~Dudal, S.~P. Sorella, N.~Vandersickel, and H.~Verschelde.
\newblock Gribov no-pole condition, zwanziger horizon function, kugo-ojima confinement criterion, boundary conditions, brst breaking and all that.
\newblock {\em Phys. Rev. D}, 84:065039, 2011.

\bibitem{Dudal2008PRDrapid}
D.~Dudal, S.~P. Sorella, N.~Vandersickel, and H.~Verschelde.
\newblock New features of the gluon and ghost propagator in the infrared region from the gribov--zwanziger approach.
\newblock {\em Physical Review D}, 77:071501, 2008.

\bibitem{Capri:2015nzw}
S.~P. Sorella et~al.
\newblock Nonperturbative aspects of euclidean yang–mills theories in linear covariant gauges: Nielsen identities and a brst-invariant two-point correlation function.
\newblock {\em Phys. Rev. D}, 92:045039, 2015.

\bibitem{Capri:2017bfd}
S.~P. Sorella et~al.
\newblock A local and brst-invariant yang–mills theory within the gribov horizon.
\newblock {\em Phys. Rev. D}, 95:045011, 2017.

\bibitem{Capri:2021pye}
R.~C. Terin et~al.
\newblock All-order renormalizable refined gribov–zwanziger model with brst-invariant fermionic horizon function in linear covariant gauges.
\newblock {\em Phys. Rev. D}, 104(5):054048, 2021.

\bibitem{Serreau:2012cg}
J.~Serreau and M.~Tissier.
\newblock Lifting the gribov ambiguity in yang-mills theories.
\newblock {\em Phys. Lett. B}, 712:97--103, 2012.

\bibitem{SerreauTissierTresmontant2015}
J.~Serreau, M.~Tissier, and A.~Tresmontant.
\newblock Covariant gauges without gribov copies.
\newblock {\em Physical Review D}, 92:105003, 2015.

\bibitem{Reinosa:2020skx}
Urko Reinosa, Julien Serreau, Rodrigo~Carmo Terin, and Matthieu Tissier.
\newblock {Symmetry restoration and the gluon mass in the Landau gauge}.
\newblock {\em SciPost Phys.}, 10(2):035, 2021.

\bibitem{Cucchieri:2008PRL}
A.~Cucchieri and T.~Mendes.
\newblock Constraints on the ir behavior of the gluon propagator in yang–mills theories.
\newblock {\em Phys. Rev. Lett.}, 100:241601, 2008.

\bibitem{DudalOliveiraSilva2018}
D.~Dudal, O.~Oliveira, and P.~J. Silva.
\newblock K{\"a}ll{\'e}n--lehmann spectroscopy for (un)physical degrees of freedom.
\newblock {\em Annals of Physics}, 397:351--364, 2018.

\bibitem{DudalOliveiraRoelfsSilva2020}
D.~Dudal, O.~Oliveira, M.~Roelfs, and P.~J. Silva.
\newblock Spectral representation for the gluon propagator and positivity violation.
\newblock {\em Nuclear Physics B}, 952:114912, 2020.

\bibitem{BinosiTripolt2020}
D.~Binosi and R.-A. Tripolt.
\newblock Spectral functions of confined particles.
\newblock {\em Physics Letters B}, 801:135171, 2020.

\bibitem{Alkofer:2001}
R.~Alkofer and L.~von Smekal.
\newblock The infrared behavior of qcd green's functions: Confinement, dynamical symmetry breaking, and hadrons as relativistic bound states.
\newblock {\em Phys. Rept.}, 353:281--465, 2001.

\bibitem{PhysRevLett.90.152001}
A.~C. Aguilar, A.~A. Natale, and P.~S.~Rodrigues da~Silva.
\newblock Relating a gluon mass scale to an infrared fixed point in pure gauge qcd.
\newblock {\em Phys. Rev. Lett.}, 90:152001, Apr 2003.

\bibitem{Huber:2020}
M.~Q. Huber.
\newblock Nonperturbative properties of yang–mills theories.
\newblock {\em Phys. Rept.}, 879:1--92, 2020.

\bibitem{Pawlowski:2007}
J.~M. Pawlowski.
\newblock Aspects of the functional renormalisation group.
\newblock {\em Annals Phys.}, 322:2831--2915, 2007.

\bibitem{Tissier:2010}
M.~Tissier and N.~Wschebor.
\newblock Infrared propagators of yang–mills theory from perturbation theory.
\newblock {\em Phys. Rev. D}, 82:101701, 2010.

\bibitem{Tissier:2011}
M.~Tissier and N.~Wschebor.
\newblock An infrared safe perturbative approach to yang–mills correlators.
\newblock {\em Phys. Rev. D}, 84:045018, 2011.

\bibitem{ParisiSourlas1979}
G.~Parisi and N.~Sourlas.
\newblock Random magnetic fields, supersymmetry and negative dimensions.
\newblock {\em Phys. Rev. Lett.}, 43:744--745, 1979.

\bibitem{PiguetSorella1995}
O.~Piguet and S.~P. Sorella.
\newblock {\em Algebraic Renormalization: Perturbative Renormalization, Symmetries and Anomalies}, volume~28 of {\em Lecture Notes in Physics Monographs}.
\newblock Springer-Verlag, Berlin, Heidelberg, 1995.

\end{thebibliography}

\end{document}